\documentclass[article, nojss]{jss}

\author{Stanislas Hubeaux \\ EPFL Lausanne \And Kaspar Rufibach \\ F. Hoffmann-La Roche AG, Basel}
\title{{\pkg{SurvRegCensCov}}: Weibull Regression for a Right-Censored Endpoint with a Censored Covariate}

\Plainauthor{Stanislas Hubeaux, Kaspar Rufibach} 
\Plaintitle{SurvRegCensCov: Weibull Regression for a Right-Censored Endpoint with a Censored Covariate} 
\Shorttitle{{\pkg{SurvRegCensCov}}: Weibull Regression with a Censored Covariate}

\Abstract{Biomarker data is often subject to limits of quantification and/or limits of detection. Statistically, this corresponds to left- or interval-censoring. To be able to associate a censored time-to-event endpoint to a biomarker covariate, the \proglang{R} package \pkg{SurvRegCensCov} provides software for Weibull regression for a right-censored endpoint, one interval-censored, and an arbitrary number of non-censored covariates. Furthermore, the package provides functions to estimate canonical parameters from censored samples based on several distributional assumptions, and a function to switch between different parametrizations used in \proglang{R} for Weibull regression. We illustrate the new software by applying it to assess Prentice's criteria for surrogacy in data simulated from a randomized clinical registration trial.}
\Plainkeywords{Weibull Regression, censored covariate, limit of detection, biomarker, surrogacy, Prentice's criteria, R} 

\Address{
  Stanislas Hubeaux \\
  \'{E}cole polytechnique f\'{e}d\'{e}rale de Lausanne \\
  E-mail: \email{stanislas.hubeaux@epfl.ch} \\

\bigskip

  Kaspar Rufibach\\
  F. Hoffmann-La Roche AG\\
  Building 670, Room 517\\
  4070 Basel, Switzerland\\
  Telephone: +41/0/61687-5341 \\
  E-mail: \email{kaspar.rufibach@roche.com}\\
  URL: \url{http://www.roche.com}
}

\usepackage{amsmath}
\usepackage{amssymb}
\usepackage{amsfonts}
\usepackage{amsthm}
\usepackage{latexsym}
\usepackage{thumbpdf}
\usepackage{color}

\newtheorem{theorem}{Theorem}[section]

\def\bea{\begin{eqnarray*}}
\def\eea{\end{eqnarray*}}
\def\be{\begin{equation}}
\def\ee{\end{equation}}
\def\bean{\begin{eqnarray}}
\def\eean{\end{eqnarray}}
\def\barr{\begin{array}}
\def\earr{\end{array}}
\def\bdes{\begin{description}}
\def\edes{\end{description}}
\def\bi{\begin{itemize}}
\def\ei{\end{itemize}}

\def\hat{\widehat}
\def\R{\mathbb{R}}

\def\Bl{\Bigl}
\def\Br{\Bigr}

\def\argmax{\mathop{\rm arg\,max}}

\newcommand{\ve}[1]{\mathbf{#1}}

\begin{document}


\section{Introduction}
\label{intro}
In survival analysis, we typically seek to analyze the association between a collection of covariates and a time-to-event endpoint. Regression models, such as Cox or parametric regression, are well-developped and available for the common situation where the endpoint is right- or even interval-censored, see e.g.,, \cite{klein_03}. However, when analyzing biomarkers it may happen that also one or more covariates are censored, typically left-censored at a limit of detection (LOD) $c$ which is given by assay sensitivity, i.e., the researcher only knows that the actual value of a biomarker, say, is $\le c$, but does not know the exact measurement. We refer to \cite{arunajadai_12} and \cite{bernhardt_14} for overviews on methods how to handle left-censored data, to \cite{langohr_04} for the description of a parametric survival regression model with a censored discrete covariate, and to \cite{armbruster_08} for a discussion of the different types of limits that appear in generation of biomarker data.


\cite{bernhardt_14} also provides a discussion of ``traditional'' approaches to treat left-censoring in covariates, namely complete case analysis or various imputation methods, e.g., replacing a left-censored observation by the limit of detection $c$ itself, by $c/2$, or $c / \sqrt{2}$. As long as the censoring does not depend on the survival time, removing observations below $c$ still yields consistent estimates, however, at the cost of reduced efficiency (\citealp[Proposition 1]{bernhardt_14}). \cite{arunajadai_12} show that na\"{i}ve implementation methods typically yield potentially hugely biased estimates and they recommend that, since there are no specific sets of conditions under which imputation methods can be guaranteed to perform well, these methods should generally not be used. In addition to their empirical inferiority, there is also not much theoretical justification for these na\"{i}ve approaches. $c/2$ could be seen as taking the mean of a Uniform random variable on $[0, c]$, but the motivation of $c / \sqrt{2}$ is not clear.

To correctly account for the nature of the data, i.e., censored time-to-event endpoint with a left-censored covariate, \cite{sattar_12} derive and maximize a likelihood function assuming a Weibull distribution for the time-to-event endpoint and one left-censored covariate. To be able to incorporate the latter in a fully parametric survival regression model, a fully-known distribution for the left-censored covariate has to be imposed. In \cite{sattar_12}, a Normal distribution is assumed for the left-censored covariate. As we are in a fully parametric model, statistical inference can be based on standard likelihood theory. So far, no publicly available software is available that provides estimates and inference in this model. \pkg{SurvRegCensCov} closes this gap. Our implementation extends the work of \cite{sattar_12} with respect to the following features:
\bi
\item We allow for estimation of the shape parameter $\gamma$ of the Weibull distribution, which was set to 1 in the simulations in \cite{sattar_12}.
\item As we outline in our example in Section~\ref{example}, in applications the incompletely observed covariate sometimes is not only left- but rather interval-censored. Our implementation allows for that data structure.
\item In the example we use for illustration, the regression model involves at least one more non-censored covariate in addition to the censored covariate. We thus extend the code described in \cite{sattar_12} to be able to take into account an arbitrary number of non-censored covariates. As a matter of fact, no non-censored covariate, i.e., a Weibull regression model involving only the censored covariate, can also be handled using \code{SurvRegCensCov}.
\item The likelihood function set up to accommodate the Weibull regression model necessitates for every censored covariate observation computation of a rather involved integral, see Section~\ref{implementation}. \cite{sattar_12} suggest to use Simpson's 1/3 rule for computation of these integrals, implying (if we understand their approach correctly) that certain tuning parameters have to be selected. We proceed differently and omit this choice of tuning parameters, see Section~\ref{implementation}.
\item Finally, we provide some functions to easily switch between different parametrizations used for Weibull regression models.
\ei

\newpage

\subsection{About this document}\label{sec:about}

The package is available from the Comprehensive \proglang{R} Archive Network at \url{http://CRAN.R-project.org/package=SurvRegCensCov} \citep{R}.

This document was created using \code{Sweave} \citep{leisch_02}, \LaTeX{} \citep{knuth_84, lamport_94}, and \proglang{R} \citep{R}. This means that all of the code has been checked by \proglang{R}.

\section{Methods}
\label{methods}

\subsection{Two-sample inference for censored Normal samples}
\cite{lynn_01} describes maximum likelihood estimation for one left-censored sample based on parametric assumptions. \code{proc lifereg} in \proglang{SAS} and the \proglang{R} package \pkg{fitdistrplus} \citep{fitdistrplus} implement that methodology for a large catalogue of distributions to choose from. \code{proc lifereg} also allows for inference based on two samples, for a variety of distributions. In the Normal case \code{proc lifereg} implicitly assumes that the variances in the two groups are identical. However, in our applications we typically do not want to entertain that restriction, and that is why we have added the function \code{NormalMeanDiffCens} to \pkg{SurvRegCensCov}. It allows for inference from two independent interval-censored samples assumed to come from Normal distributions with different means and variances. Note that the default setting for the \code{t.test} function, i.e., inference from two completely observed Normal samples, is indeed as well \code{var.equal = FALSE}. We illustrate the application of \code{NormalMeanDiffCens} in Section~\ref{example}.

%
%
%
%


In addition to a function for the two-sample Normal case, \pkg{SurvRegCensCov} also provides a function (\code{ParamSampleCens}) to estimate canonical parameters from one sample, for the Normal, Logistic, Gamma, and Weibull distribution. Note that the function \pkg{fitdistcens} in the package \pkg{fitdistrplus} \citep{fitdistrplus} offers similar functionality.

\subsection{Weibull regression model with a censored covariate - the model}
Define a triple $(T, \delta, \ve{X})$, where $T$ denotes the follow-up time, $\delta$ is the censoring indicator, and $\ve{X} \in \mathbb{R}^d$ is a (time-invariant) vector of baseline covariates. The actual survival time is $Z$ and the censoring time $C$, so that we observe $T = \min\{Z, C\}$ and $\delta = 1\{T \le Z\}$, with $1\{ \cdot \}$ the indicator function. The survival time $Z$ is assumed to be conditionally independent given $\ve{X}$, i.e., the censoring is supposed to be non-informative. \cite{sattar_12} then assume that $Z$ follows a Weibull distribution with shape parameter $\gamma$ and scale parameter $\lambda$, so that the density and hazard function of $Z$ are defined as
\bean
  f_0(z) \ = \ \lambda \gamma z^{\gamma - 1} \exp(-\lambda z^\gamma) && h_0(z)\ = \ \lambda \gamma z^{\gamma - 1}.  \label{param}
\eean This is the parametrization used by \citet[Chapter 12]{klein_03}. Note that \proglang{R}'s \code{dweibull} function uses another parametrization, namely
\bea
  f^*_0(z) \ = \ (a/b) (z/b)^{a - 1} \exp\Bl(-(z/b)^a\Br) && h^*_0(z)\ = \ (a/b) (z/b)^{a - 1},
\eea so that the reparametrization is $\gamma = a, \lambda = b^{-a}$. Yet another parametrization is used by the \code{survreg} function in \pkg{survival} \citep{survival-package, survival-book}, see \eqref{repar survreg}.

Using Parametrization~\eqref{param} to set up a Weibull regression, we then assume that the hazard function, for a given covariate vector $\ve{X}$ and a corresponding vector $\ve{\beta}$ of regression parameters, can be written as
\bea
  h(z | \ve{X}) &=& \exp(\ve{\beta}^\top \ve{X}) h_0(z) \\
                &=& \exp(\ve{\beta}^\top \ve{X}) \lambda \gamma z^{\gamma - 1} \\
                &=:& \lambda^\star \gamma z^{\gamma - 1}.
\eea
The last equality shows that we model the scale parameter $\lambda^\star = \lambda \exp(\ve{\beta}^\top \ve{X})$ using a base scale $\lambda$ and covariates. The coefficients $\exp\ve{\beta}$ feature the proportional hazards property, i.e., can be interpreted as hazard ratios.
On the other hand, \code{survreg} embeds Weibull regression in the framework of a more general accelerated failure time model and its output provides estimates for $\log(\sigma)$ (denoted \code{Log(scale)}), $-\mu / \sigma$ (\code{Intercept}), and the regression parameters $\alpha$. The reparametrization from \eqref{param} is
\bean
  \gamma \ = \ \sigma^{-1}, \ \ \lambda \ = \ \exp(-\mu / \sigma), \ \ \ve{\beta} \ = \ - \alpha / \sigma. \label{repar survreg}
\eean
In \pkg{SurvRegCensCov}, the function \code{ConvertWeibull} takes the output of \code{survreg} as input and transforms parameter estimates to Parametrization \eqref{param}, thereby allowing to easily switch from the accelerated failure time to the proportional hazard interpretation. Transformed confidence intervals, based on applying the $\delta$-rule to \code{survreg}'s standard errors (see \citealp{klein_03}, Section 12.2), are available as well. \code{ConvertWeibull} also computes the ``acceleration factor'' $\exp \ve{\alpha}$, alternatively called ``event time ratio'', see e.g., \cite{carroll_03} for how to interpret it. Note that the Weibull distribution is the only parametric family that allows to set up a regression model that allows for both, a proportional hazard and an accelerated failure time interpretation.

\subsection{Maximum likelihood estimation of parameters}
\label{sec: MLest}
Now assume we have an i.i.d. sample $(T_i, \delta_i, \ve{X}_i), i = 1, \ldots, n$ of observations and want to estimate the parameter vector $(\gamma, \lambda, \ve{\beta})$ based on this data and maximum likelihood. The corresponding likelihood function for completely observed $\ve{X}_i$ is, according to \cite{sattar_12},
\bea
  L_1(\gamma, \lambda, \ve{\beta}) &=& \prod_{i=1}^n h(T_i | \ve{X})^{\delta_i} S(T_i | \ve{X})\\
  &=& \prod_{i=1}^n f(T_i | \ve{X})^{\delta_i} S(T_i | \ve{X})^{1-\delta_i}.
\eea
For the moment, let us further assume that we observe some of the entries of the first covariate $(X_{11}, \ldots, X_{1n})$ only if it is above the patient-specific LOD $c_i$ and that for those values below $c_i$, we simply use some imputation scheme, typically just impute $c_i$. Then, maximum likelihood estimates (MLE) based on $L_1$ are in general biased \citep{dangelo_08, sattar_12}. In order to account for $X_{1i}$ potentially being censored, \cite{sattar_12} make additional assumptions and modify the likelihood function as follows: First, we explicitly specify a distribution for $X_{1i}$ by specifying the density $f_\theta$, where $\theta \in \R^d$ indicates the parametrization of $f$, for $d \ge 1$. Note that they assume the same distribution for all observations. Often, it seems sensible to assume $f_\theta = f_{\mu, \sigma^2}$ as Normal, maybe after taking the logarithm of $X_{1i}$. Define for each observation the binary random variable $R_i := 1\{X_{1i} \ge c_i\}$ that indicates the status of the observation, i.e., whether it is observed or left-censored. The probability mass function $p_{R_i}$ of $R_i$ is a simple Bernoulli distribution with success probability $\pi_i = P(X_{1i} > c_i) = \int_{c_i}^\infty f_\theta(x) d x$ that $X_{1i}$ is observed.

\citet[Chapter 3.5]{klein_03} explain how arbitrarily censored data contributes to the likelihood function when we estimate parameters from a random sample. \citet[Formula 2]{sattar_12} use this to construct the Weibull regression likelihood function based on our available data including a left-censored covariate, yielding
\bea
  L_2(\gamma, \lambda, \ve{\beta}) &=& \prod_{i=1}^n \Bl( f(T_i | \ve{X})^{\delta_i} S(T_i | \ve{X})^{1-\delta_i} p_{R_i}(r_i) f_\theta(X_{1i}) \Br)^{r_i} \times \\
  && \hspace{0.5cm} \Bl( \int_{-\infty}^{c_i} f(T_i | \ve{X})^{\delta_i} S(T_i | \ve{X})^{1-\delta_i} p_{R_i}(r_i) f_\theta(x) d x \Br)^{1 - r_i}.
\eea This likelihood can in general not be written in closed form and has thus to be maximized numerically. The MLE is defined as
\bea
  (\hat \gamma, \hat \lambda, \ve{\hat \beta}) &:=& \argmax_{(\gamma, \lambda, \ve{\beta}) \in \mathbb{R}^{2 + d}} \log L_2(\gamma, \lambda, \ve{\beta}).
\eea
Extension to interval-censoring and more than one censored covariate is straightforward, but tedious. The function \code{SurvRegCov} in \pkg{SurvRegCensCov} provides standard errors of the estimates $(\hat \gamma, \hat \lambda, \ve{\hat \beta})$ based on the observed Fisher information matrix.

\section{Implementation}
\label{implementation}
To maximize $L_2$, we use \proglang{R}'s function \code{optim}. Note that maximization of $L_2$ is not entirely straightforward, as for each evaluation of $L_2$ at a given value of the parameter vector $(\gamma, \lambda, \ve{\beta})$, $n - \sum_{i=1}^{n} R_i$ integrals have to be numerically computed. As discussed in Section~\ref{intro}, \cite{sattar_12} use an approximation rule to compute the integrals, and - if we interpret their description of implementation correctly - their approach necessitates the choice of the tuning parameters $a, b, h$, where $[a, b]$ is the interval the function to be integrated ``lives on'' and $h$ is the grid length. In our initial implementation of \cite{sattar_12}'s approach, the choice of these parameters might depend on the actual data, i.e., the maximization of the log-likelihood function potentially needs to be tuned manually for a given dataset. Instead, we simply set very small values of the function to be integrated to 0. We can then use \proglang{R}'s \code{integrate} function to do the repeated integrations, implying a considerable gain in computational speed without sacrifying numerical accuracy of the resulting estimates. The modification also implies that the maximization of the log-likelihood function in our \pkg{SurvRegCensCov} is fully automatic and does not necessitate the choice of tuning parameters.

To compute the Hessian matrix of the log-likelihood function we use the package \pkg{numDeriv} \citep{numDeriv}. The latter generally provides more accurate computations of derivatives in complex problems than \code{optim}.

The function \code{SurvRegCens} in \pkg{SurvRegCensCov} implements maximization of $L_2$, and provides standard statistical inference for the resulting parameter estimate $(\hat \gamma, \hat \lambda, \ve{\hat \beta})$. In addition to this main function of the package, \pkg{SurvRegCensCov} provides \code{ConvertWeibull} to reparametrize the output of \code{survreg}, \code{WeibullReg} does these two steps in one function, and \code{WeibullDiag} constructs a diagnostic plot to visually check the adequacy of the Weibull distribution for survival data with respect to one nominal covariate.

\section{Examples}
\label{example}

\cite{sattar_12} illustrate Weibull regression with a left-censored covariate and no completely observed covariate using the Genetic and Inflammatory Marker of Sepsis (GenIMS) study \citep{yende_08}. The interest was to associate the right-censored endpoint to some cytokine levels which were subject to a uniform LOD $c$.

\subsection{Surrogacy}
\label{surrogacy}
Our incentive to implement the methodology came from another application: So far, registrational studies for new drugs in chronic lymphoid leukemia (CLL) typically used progression-free survival (PFS) as their primary endpoint. With the advent of ever more potent therapies in this disease and luckily for patients, PFS is getting longer and longer, especially in those patients that are considered ``fit'' for aggressive chemotherapy. As an example, the patients in the treatment arm of the CLL8 study \citep{hallek_10}, a randomized trial in first-line CLL patients, were seen with a median PFS of 57 months \citep{fischer_12}. As this improvement in PFS continues, clinical studies with PFS as an endpoint become infeasible and surrogate endpoints for PFS are sought. In addition, several treatment regimens are entering the CLL landscape which tackle the disease from different angels and are or will be approved in the near future. In order to establish efficacy of these new therapies, or even combinations between them, ``early'' measurable endpoints that replace the ever growing PFS are needed. Such a potential early measurable and surrogacy candidate is minimal residual disease (MRD, \citealp{bottcher_12}), a continuous endpoint that basically counts the number of cancerous cells left in the body after therapy. MRD can be measured either by IGH polymerase chain reaction (PCR) or flow cytometry. The latter method was used in CLL8 and exhibited a sensitivity of $c := 10^{-4}$, i.e., some of the measurements are indeed left-censored at $c$.

Note that MRD is not a biomarker in the strict sense, but rather an endpoint that measures effect of therapy after patient has undergone treatment. However, left-censoring often occurs when analyzing biomarker data, so that our new software is applicable in a much broader sense than to assess surrogacy only.

The CLL11 study \citep{goede_14} established superiority of a 2$^\text{nd}$ generation CD20 antibody, obinutuzumab, over rituximab, a 1$^\text{st}$ generation antibody. In this trial, MRD was measured using PCR via a patient-specific assay. Furthermore, as is typical for small MRD values, the exact value can often not be measured, so that the labs provided a lower and upper limit of detection in these cases, i.e., some of the measurements were in fact genuinely interval-censored. As a consequence, in order to establish surrogacy based on CLL11 data, we indeed needed to extend the setup in \cite{sattar_12} to allow for interval-censored data.

To formally establish surrogacy, the first step is typically to assess what is called ``Prentice's criteria'' \citep{prentice_89}. The goal is to show that the hazard for PFS is independent of treatment, conditional on the surrogate variable MRD. In order to do so, a regression model with PFS as endpoint and MRD as well as treatment as covariates has to be set up. Prentice's criteria can be considered met if in such a model, MRD is associated with the endpoint and captures all the effect of the treatment, i.e., the latter is not anymore associated to the endpoint. Thus, Prentice's criteria is typically regarded as fulfilled if treatment is not significant anymore at a pre-specififed significance level in this regression model and has a small effect estimate.

In \cite{bottcher_12}, MRD was discretized in three categories using cutoffs $10^{-4}$ and $10^{-2}$. Discretization might make sense clinically, but from a statistical point of view it is well-known that by discretization we loose up to 1/3 of the information in a given continuous variable \citep{lagakos_88}. So one of our goals with implementing \pkg{SurvRegCensCov} was to assess Prentice's criteria for the CLL11 study using the full MRD information, i.e., considering the latter a continuous - but interval-censored - covariate in a Weibull regression model.

\subsection{Imitating the CLL11 study}
\label{cll11}
\cite{goede_14} is the first paper on CLL11 that reports clinical results and only contains very high-level MRD results. No detailed analysis of MRD in CLL11 has been published yet and we are thus not able to illustrate \pkg{SurvRegCensCov} using original CLL11 data, although this study actually initiated our work and motivated the extensions discussed above.

To understand the features of our new and two potential traditional approaches in the context of CLL11, we instead simulate data with similar properties than the one in the rituximab (R, $n = 226$) and obinutuzumab (O, $n = 221$) treatment arms in CLL11 to illustrate \pkg{SurvRegCensCov}. To this end, we first explored the MRD data. As is typical for biomarker data, original MRD measurements were right-skewed, so that we used $\log$(MRD) in our model. The transformed measurements nicely followed a Normal distribution. Since $\log$(MRD) is a continuous interval-censored variable, we used \code{ParamSampleCens} to estimate $\mu$ and $\sigma$ of the assumed Normal distribution, in each treatment arm separately. Applying the function \code{SurvRegCens} we finally estimated the canonical parameters $\lambda, \gamma$ and the regression coefficients $\beta_\text{tmt}$, $\beta_\text{MRD}$ corresponding to treatment and $\log$(MRD) in our Weibull regression model. We end up with eight estimated parameters $\hat \mu_\text{R}, \hat \sigma_\text{R}, \hat \mu_\text{O}, \hat \sigma_\text{O}, \hat \lambda, \hat \gamma, \hat \beta_\text{tmt}$ and $\hat \beta_\text{MRD}$ based on CLL11 data. In addition, we also determined the proportion of censored MRD observations in each treatment arm. Using these estimated parameters and the assumptions of Normality of the density $f_\theta$ of the censored covariate and Weibull distributed PFS times, we can simulate right-censored PFS times. Contrasting these simulated PFS times to the original CLL11 data, we get strikingly similar Kaplan-Meier estimates.

To mimick CLL11 data (which we are not able to show at this point) in a small simulation study, we thus sufficiently varied the eight parameters above that define our model, so that no conclusion on the true nature of CLL11 data is possible. Specifically, we assumed $\mu_\text{R} = -1.5, \sigma_\text{R} = 1.5, \mu_\text{O} = -3.5, \sigma_\text{O} = 1.5, \lambda = 0.75, \gamma = 3.1, \beta_\text{tmt} = 0$ and $ \beta_\text{MRD} = 0.7$ as well as proportions of 0.05 and 0.35 of left-censored MRD values as well as a sample size of $n = 200$ in each simulated arm. We note that the higher proportion of left-censored MRD values in the obinutuzumab arm is indicative of the higher efficacy of that treatment. We deliberately choose the model to have no treatment effect, as this corresponds to the ``ideal'' situation when indeed MRD is a complete surrogate for PFS. Since in our simulation study we compare Weibull regression with a censored covariate to Weibull and Cox regression where we impute the censored observations by the LOD, we only left-censor MRD in our simulation. This implies that we can compare the newly implemented to the latter more standard methods with imputation at the LOD. MRD values were left-censored at the quantiles corresponding to the empirical censoring proportions $0.05$ and $0.35$ in CLL11.

\begin{figure}[t!]
\begin{center}
\setkeys{Gin}{width=\textwidth}
\includegraphics{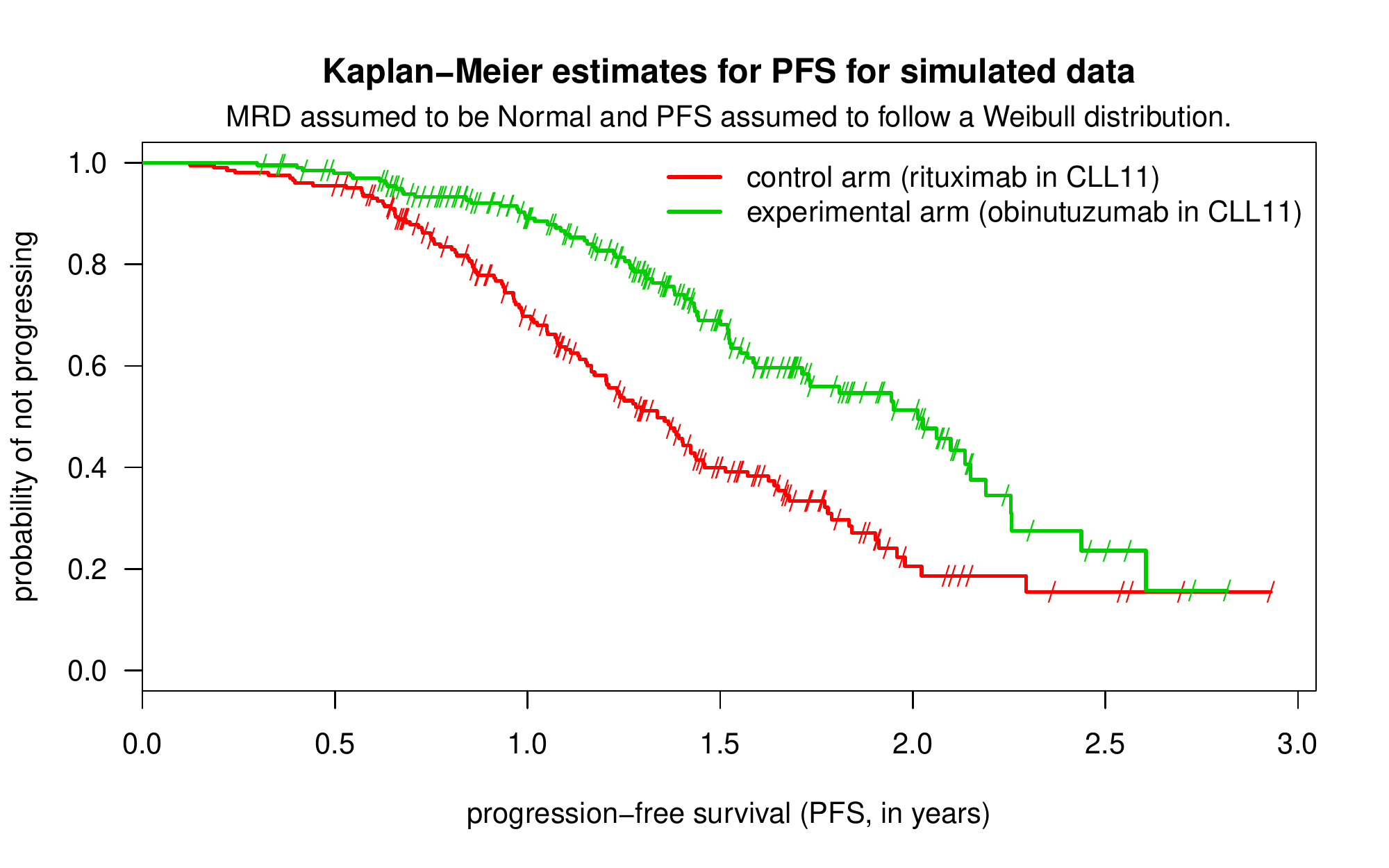}
\end{center}
\vspace*{-1cm}
\caption{Kaplan-Meier estimates of PFS for first simulated sample.}
\label{km1}
\end{figure}

To illustrate \code{SurvRegCens} the chunk below provides its model output received from estimating the model based on the first simulated sample and Figure~\ref{km1} provides the Kaplan-Meier estimates of this simulated sample:

\begin{Schunk}
\begin{Sinput}
R> WRmod1 <- SurvRegCens(formula = formula(Surv(time, event) ~
+                 Surv(time = mrd.low, time2 = mrd.up, type = "interval2") +
+                 tmt.df), data = data.frame(time, event, mrd.low, mrd.up,
+                 tmt.df), Density = DensityCens, initial = initial,
+                 namCens = "mrd", trace = 0)
\end{Sinput}
\end{Schunk}
\begin{Schunk}
\begin{Sinput}
R> summary(WRmod1)
\end{Sinput}
\begin{Soutput}
Weibull regression for a right-censored response with an interval-censored covariate

Call:
SurvRegCens(formula = formula(Surv(time, event) ~ Surv(time = mrd.low,
    time2 = mrd.up, type = "interval2") + tmt), data = data.frame(time,
    event, mrd.low, mrd.up, tmt), Density = DensityCens, initial = initial,
    namCens = "mrd", trace = 0)

Coefficients:
       Estimate Std. Error   CI.low    CI.up p-value
lambda  0.74595    0.08455  0.58024  0.91166      NA
gamma   2.98008    0.16942  2.64802  3.31213      NA
tmt     0.14092    0.16852 -0.18937  0.47121   0.403
mrd     0.67344    0.06010  0.55565  0.79122  <2e-16

AIC: 2290.1
\end{Soutput}
\end{Schunk}

\newpage

The model is well able to estimate the parameters of interest and the estimated coefficient of treatment is indeed close to 0, or the hazard ratio \code{exp(Estimator)} close to 1, and non-significant at $\alpha = 0.05$. The MRD covariate is specified as interval-censored using the coding for \code{type = interval2} in \code{Surv} \citep{survival-package}:

\begin{Schunk}
\begin{Sinput}
R> head(data.frame(mrd.low, mrd.up))
\end{Sinput}
\begin{Soutput}
  mrd.low  mrd.up
1 -1.5124 -1.5124
2 -0.7509 -0.7509
3 -1.7824 -1.7824
4 -1.9962 -1.9962
5 -2.6862 -2.6862
6      NA -3.9673
\end{Soutput}
\end{Schunk}

The estimated parameters of $\log$(MRD) in each treatment arm are
\begin{Schunk}
\begin{Sinput}
R> NormalMeanDiffCens(censdata1 = mrd.r, censdata2 = mrd.o)
\end{Sinput}
\begin{Soutput}
                      Estimator Std. Error CI.low  CI.up p-value
mu1                      -1.532    0.10429 -1.737 -1.328  <2e-16
mu2                      -3.358    0.10007 -3.554 -3.162  <2e-16
sigma1                    1.468    0.07707  1.317  1.619 1.2e-09
sigma2                    1.329    0.08474  1.163  1.496   1e-04
Mean difference delta     1.826    0.14454  1.542  2.109  <2e-16
\end{Soutput}
\end{Schunk}

Again, the underlying true parameters are accurately estimated. However, note that in our model in Section~\ref{sec: MLest} we assume the same density $f_\theta$ for the censored covariate for all observations, so that to estimate \code{WRmod1} above we used \code{DensityCens} defined according to
\begin{Schunk}
\begin{Sinput}
R> res1 <- ParamSampleCens(censdata = rbind(mrd.r, mrd.o), dist = "normal")
R> res1$coeff
\end{Sinput}
\begin{Soutput}
      Estimator Std. Error CI.low  CI.up p-value
Mu       -2.467    0.08774 -2.639 -2.295  <2e-16
Sigma     1.712    0.06975  1.576  1.849  <2e-16
\end{Soutput}
\begin{Sinput}
R> DensityCens <- function(value, res = res1){return(dnorm(value,
+                          mean = res$coeff[1, 1] , sd = res$coeff[2, 1]))}
\end{Sinput}
\end{Schunk}

\subsection{Simulation results}
\label{simulations}
\begin{figure}[t!]
\begin{center}
\setkeys{Gin}{width=\textwidth}
\includegraphics{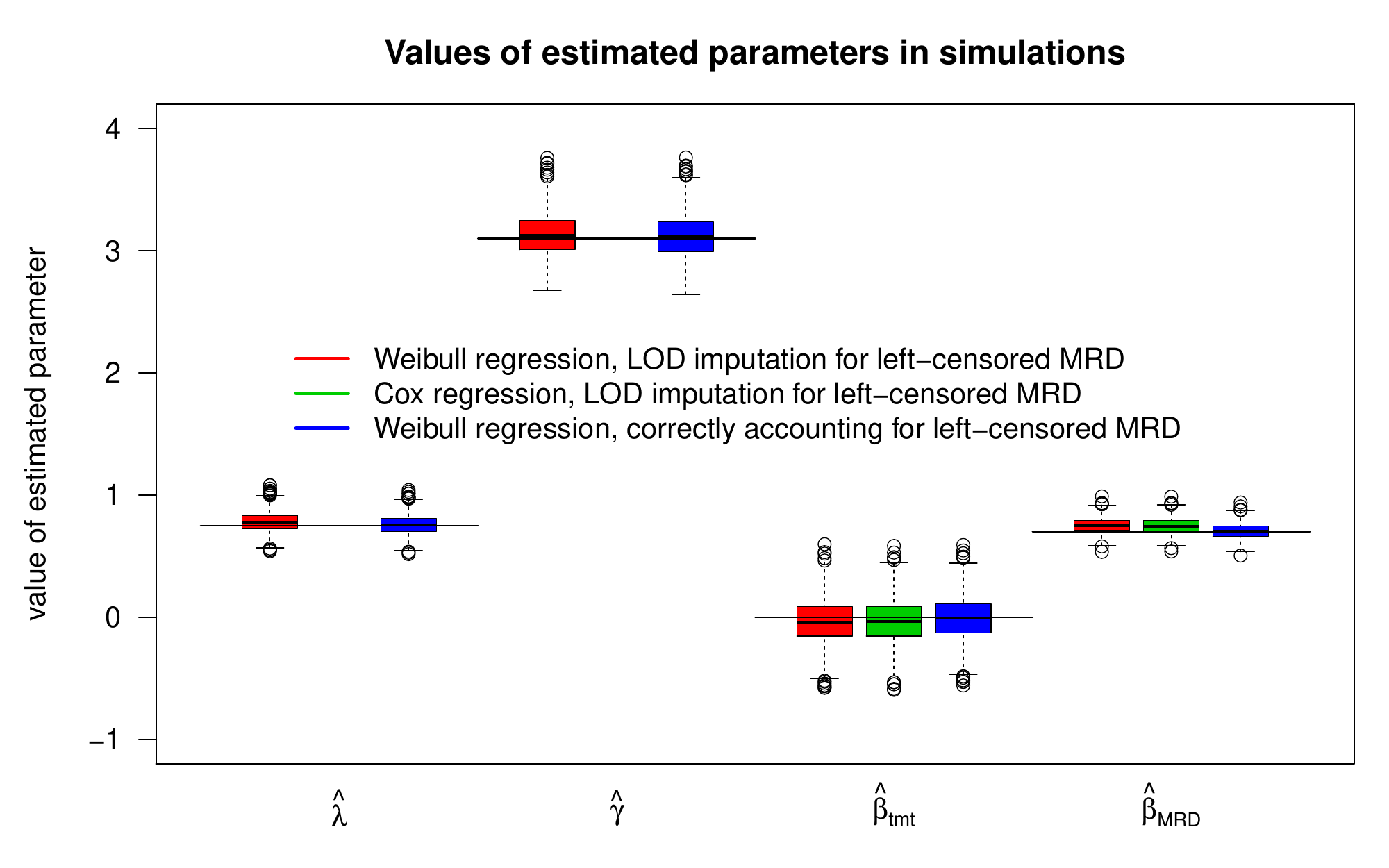}
\end{center}
\vspace*{-1cm}
\caption{Boxplots of estimated parameters in 1000 simulations.}
\label{param est box}
\end{figure}

We generated $M = 1000$ ``CLL11 like'' trials using the parameters above and for each simulation we estimated three regression models for PFS and explanatory variables treatment and $\log$(MRD):
\begin{itemize}
\item Weibull regression using the function \code{survreg} in \pkg{survival}, where left-censored MRD values were considered to be actual measurements, i.e., we impute the lower limit of detection as an actual MRD value.
\item Cox regression with the same imputation strategy, using the function \code{coxph} in \pkg{survival}.
\item Weibull regression assuming normally distributed $\log$(MRD), correctly accounting for the censoring of MRD via the function \code{SurvRegCensCov}.
\end{itemize}
As discussed in Section~\ref{intro}, imputing the limit of detection for left-censord values is common in applications. Figure~\ref{param est box} provides boxplots of the $1000$ estimated parameters whereas Table~\ref{bias} displays the bias of each method in estimating the Weibull and/or regression parameters.

\newpage

\begin{table}[h]
\centering
{\footnotesize
\begin{tabular}{lrrrr}
  \hline
 & $\hat \lambda$ & $\hat \gamma$ & $\hat \beta_\text{tmt}$ & $\hat \beta_\text{MRD}$ \\
  \hline
True parameter value we simulated from & 0.75 & 3.1 & 0 & 0.7 \\
   \hline
Bias Weibull regression, correctly accounting for left-censored MRD &  0.0069 &  0.0231 & -0.0060 &  0.0046 \\
  Bias Weibull regression, LOD imputation for left-censored MRD &  4.7 &  5.6 &  4.8 &  8.5 \\
  Bias Cox regression, LOD imputation for left-censored MRD &  &  &  1.3 & 10.3 \\
   \hline
\end{tabular}
}
\caption{Bias for Weibull regression with censored covariate and relative bias for imputation methods. Rows 2 and 3 are factors by which Row 1 is multiplied to get the bias of the other methods. Biases for $\hat \beta_\text{tmt}$ were negative for all three methods.}
\label{bias}
\end{table}
We find that absolute bias is moderate for all three methods, but that clearly Weibull regression correctly accounting for the left-censoring of MRD performs best. This remains true when we look at the mean-squared error in Table~\ref{mse}. Finally, in an analysis of surrogacy we are interested in testing the null hypothesis of no treatment effect in a hypothesis test. As we simulated assuming that $\beta_\text{tmt} = 0$, by counting how often in our simulations a test for this hypothesis at a significance level of $\alpha = 0.05$ rejects we get an empirical assessment how well each method keeps the probability of a type I error. Weibull regression correctly accounting for left-censoring yielded a proportion of 0.05 rejected null hypothesis for the treatment regression parameter, Weibull regression with LOD imputation 0.062, and Cox regression with LOD imputation 0.059. Thus the maximum likelihood estimate exactly keeps the probability of a type I error, whereas the two approximate methods slightly inflate it.

\begin{table}[h]
\centering
{\footnotesize
\begin{tabular}{lrrrr}
  \hline
 & $\hat \lambda$ & $\hat \gamma$ & $\hat \beta_\text{tmt}$ & $\hat \beta_\text{MRD}$ \\
  \hline
True parameter value we simulated from & 0.75 & 3.1 & 0 & 0.7 \\
   \hline
MSE Weibull regression, correctly accounting for left-censored MRD & 0.00005 & 0.00053 & 0.00004 & 0.00002 \\
  MSE Weibull regression, LOD imputation for left-censored MRD &  21.9 &  31.4 &  22.8 &  72.7 \\
  MSE Cox regression, LOD imputation for left-censored MRD &  &  &   1.8 & 105.9 \\
   \hline
\end{tabular}
}
\caption{Mean-squared error for Weibull regression with censored covariate and relative bias for imputation methods. Rows 2 and 3 are factors by which Row 1 is multiplied to get the MSE of the other methods.}
\label{mse}
\end{table}

\section{Discussion}
\label{discussion}
Biomarker measurements with lower and/or upper limit of detection are common in applied statistics. Motivated by a concrete application in a large clinical registration study, namely to establish surrogacy of MRD for PFS in CLL, we implemented an extension to the method initially proposed in \cite{sattar_12} and collected the code in the package \code{SurvRegCensCov}. \cite{bernhardt_14} note that maximizing $L_2(\gamma, \lambda, \beta)$ can be extremely slow, due to the multiple integrals that need to be evaluated. Computation of the maximizer is indeed not immediate, but still sufficiently efficient in a typical application.

In our simulation, assuming that proportions of $0.05$ and $0.35$ of observations of the censored covariate were actually left-censored, we found that \code{SurvRegCensCov} delivers estimates with lower bias and MSE compared to the traditional methods. The latter also slightly inflated the probability of a type I error for a hypothesis test on the regression parameter for treatment. However, bias for the traditional methods was still moderate. Nevertheless, the advantage of the MLE increases once more values of $X_1$ are in fact censored, once they are also interval- and not only left-censored, and in case we analyze less observations.

Potential extensions of our implementation are observation-specific assumptions for $f_\theta$: In our example, that would enable to e.g., assume different densities for the censored covariate in each treatment group. However, as the bias of estimated parameters is already very small with the assumption of the same density for all observations, we do not anticipate a large gain by that generalization in typical setups. Other parametric assumptions than Weibull for $Z$ and Normal for $X_1$ are straightforward to implement. \code{SurvRegCensCov} allows for one censored covariate only. Methodologically, extension to more censored covariates is straightforward, however, the number of integrals to be computed to evaluate the corresponding maximum likelihood function would substantially increase, slowing down computations.

\section*{Acknowledgments}
This paper summarizes work that Stanislas Hubeaux has done as an intern in the Biostatistics Oncology Department at F. Hoffmann-La Roche AG in Basel, in collaboration with Kaspar Rufibach.

We thank Elina Asikanius, Paul Delmar, and Carol Ward for many helpful discussions and proofreading the manuscript.

The functions \code{ConvertWeibull}, \code{WeibullReg}, and \code{WeibullDiag} in \pkg{SurvRegCensCov} have been written by Sarah Haile and we thank her for allowing us to include them in \pkg{SurvRegCensCov}.


\bibliography{stat}

\end{document}